\documentstyle[pmart,twoside,fleqn,epsf]{article} 

\begin{document}            

%
%
\title{SPIN-$\frac{1}{2}$ PERIODIC NONUNIFORM $XX$ CHAINS
       AND THE SPIN-PEIERLS INSTABILITY}

\author{O. Derzhko$^{a,b}$, J. Richter$^{c}$ and O. Zaburannyi$^{b}$}

%
%
\address{$^{a}$Institute for Condensed Matter Physics,\\
               1 Svientsitskii St., L'viv-11, 290011, Ukraine\\
         $^{b}$Chair of Theoretical Physics,
               Ivan Franko State University of L'viv,\\
               12 Drahomanov St., L'viv-5, 290005, Ukraine\\
         $^{c}$Institut f\"{u}r Theoretische Physik,
               Universit\"{a}t Magdeburg,\\
               P.O. Box 4120, D-39016 Magdeburg, Germany}
%
%

\date{April 27, 1999}

\maketitle                   
%
%

\begin{abstract}
  Using continued fractions we obtained the exact result
  for the density of magnon states
  of the regularly alternating spin-$\frac{1}{2}$ $XX$ chain
  with Dzyaloshinskii-Moriya interaction.
  We examined the stability of the magnetic chain
  with respect to the spin-Peierls dimerization.
\end{abstract}

%
%

Since the discovery of the inorganic spin-Peierls compound
CuGeO$_3$
the interest in the properties of spin-Peierls systems
considerably increased
[1].
The models that can be examined exactly play an important role
in clarifying the generic features of such systems.
An example of such a model is the spin-$\frac{1}{2}$ $XX$ chain
that was studied in several papers
[2, 3, 4]
(note, however, that in the
non-adiabatic limit
such a spin chain does not permit
exact analysis
[5]).
The aim of the present study is to examine the influence
of an additional Dzyaloshinskii-Moriya coupling
on the
spin-Peierls dimerization.
The presence of such a term for CuGeO$_3$ was proposed
in
[6, 7].
The multisublattice spin-$\frac{1}{2}$ $XX$ chain
with the Dzyaloshinskii-Moriya interaction was introduced in
[8].
In our study we follow the idea of
[2]
and compare the total ground state energy
of the dimerized and uniform chains.
However, in contrast to previous works
[2, 3, 4, 8]
we use the continued-fraction
representation for the one-fermion Green functions
[9]
that allows
a natural extension
of the calculations for more complicated lattice distortions
having finite period.

We consider $N\rightarrow \infty$ spins $\frac{1}{2}$ on a circle
with the Hamiltonian
\begin{eqnarray}
H=\sum_{n}\Omega_ns_n^z
+2\sum_{n}
I_n\left(s_n^xs_{n+1}^x+s_n^ys_{n+1}^y\right)
\nonumber\\
+2\sum_{n}
D_n\left(s_n^xs_{n+1}^y-s_n^ys_{n+1}^x\right).
\end{eqnarray}
After the Jordan-Wigner transformation one comes to tight-binding
spinless fer\-mi\-ons
on a circle with complex hopping integrals.
We introduce the temperature double-time
one-fermion Green functions that yield the density of magnon states
$\rho(E)
=\mp\frac{1}{\pi N}\sum_n{\mbox{Im}}G_{nn}^{\mp}$,
$G_{nm}^{\mp}\equiv G_{nm}^{\mp}(E\pm i\epsilon)$.
We further make
use of the continued-fraction representation for the required
diagonal Green functions
\begin{eqnarray}
G_{nn}^{\mp}=\frac{1}{E\pm i\epsilon-\Omega_n-\Delta_n^--\Delta_n^+},
\nonumber\\
\Delta_n^-=\frac{I^2_{n-1}+D^2_{n-1}}
{E\pm i\epsilon -\Omega_{n-1}-\frac{I^2_{n-2}+D^2_{n-2}}
{E\pm i\epsilon -\Omega_{n-2}-_{\ddots}}},
\nonumber\\
\Delta_n^+=\frac{I^2_{n}+D^2_{n}}
{E\pm i\epsilon -\Omega_{n+1}-\frac{I^2_{n+1}+D^2_{n+1}}
{E\pm i\epsilon -\Omega_{n+2}-_{\ddots}}}.
\end{eqnarray}
One immediately notes that for any finite period of varying
$\Omega_n$, $I_n$, $D_n$ the continued fractions
$\Delta^-_n$, $\Delta^+_n$
involved into $G^{\mp}_{nn}$ (2)
become periodic
and thus can be calculated exactly
yielding the exact result for the density of states
and hence for the thermodynamic quantities of spin model (1).
For example, for the periodic chain having period 2
$\Omega_1I_1D_1\Omega_2I_2D_2\Omega_1I_1D_1\Omega_2I_2D_2\ldots $
the described scheme gives
\begin{eqnarray}
\rho(E)=
\left\{
\begin{array}{ll}
0, &
{\mbox{if}}\;\;\;E\le b_4,\;b_3\le E\le b_2,\;b_1\le E,\\
\frac{1}{2\pi}\frac{\vert 2E-\Omega_1-\Omega_2\vert}
{\sqrt{{\cal{B}}(E)}},&
{\mbox{if}}\;\;\;b_4<E<b_3,\;b_2<E<b_1,
\end{array}
\right.
\nonumber\\
{\cal{B}}(E)
=4{\cal{I}}_1^2{\cal{I}}_2^2
-\left[
(E-\Omega_1)(E-\Omega_2)-{\cal{I}}_1^2-{\cal{I}}_2^2
\right]^2
\nonumber\\
=-(E-b_4)(E-b_3)(E-b_2)(E-b_1),
\nonumber\\
\left\{b_4\le b_3\le b_2\le b_1\right\}
=
\left\{
\frac{1}{2}(\Omega_1+\Omega_2)\pm {\sf{b}}_1,
\frac{1}{2}(\Omega_1+\Omega_2)\pm {\sf{b}}_2
\right\},
\nonumber\\
{\sf{b}}_1=\frac{1}{2}
\sqrt{\left(\Omega_1-\Omega_2\right)^2
+4\left(\vert{\cal{I}}_1\vert+\vert{\cal{I}}_2\vert\right)^2},
\nonumber\\
{\sf{b}}_2=\frac{1}{2}
\sqrt{\left(\Omega_1-\Omega_2\right)^2
+4\left(\vert{\cal{I}}_1\vert-\vert{\cal{I}}_2\vert\right)^2},
\end{eqnarray}
where ${\cal{I}}^2_n=I_n^2+D_n^2$.

To examine the instability
of the considered spin chain
with respect to dimerization we may assume
$\vert I_1\vert=\vert I\vert(1+\delta)$,
$\vert D_1\vert=\vert D\vert(1+\delta)$,
$\vert I_2\vert=\vert I\vert(1-\delta)$,
$\vert D_2\vert=\vert D\vert(1-\delta)$,
$0\le\delta\le 1$
restricting ourselves to a case of the uniform transverse field
$\Omega_1=\Omega_2=\Omega_0$.
The total energy per site
${\cal{E}}(\delta)$
consists of the magnetic part
$e_0(\delta)$
that follows from (3)
\begin{eqnarray}
e_0(\delta)=-\frac{1}{2}\int_{-\infty}^{\infty}
dE\rho(E)\vert E\vert
=-\frac{2\vert{\cal{I}}\vert}{\pi}{\mbox{E}}(\psi,1-\delta^2)
-\vert\Omega_0\vert
\left(
\frac{1}{2}-\frac{\psi}{\pi}
\right),
\end{eqnarray}
where the effective interspin coupling
$\vert{\cal{I}}\vert=\sqrt{I^2+D^2}$ has been introduced,
\linebreak
${\mbox{E}}(\psi,a^2)\equiv
\int_0^{\psi}d\phi\sqrt{1-a^2\sin^2\phi}$
is the elliptic integral of the second kind,
$\psi=0$ if $2\vert{\cal{I}}\vert\le\vert\Omega_0\vert$,
$\psi={\mbox{arcsin}}\sqrt{\frac{4{\cal{I}}^2-\Omega_0^2}
{4{\cal{I}}^2(1-\delta^2)}}$
if $2\delta\vert{\cal{I}}\vert\le\vert\Omega_0\vert<2\vert{\cal{I}}\vert$,
$\psi=\frac{\pi}{2}$ if $\vert\Omega_0\vert<2\delta\vert{\cal{I}}\vert$,
and the elastic part $\alpha\delta^2$.
Besides the trivial solution
$\delta^{\star}=0$
the equation
$\frac{\partial{\cal{E}}(\delta)}{\partial\delta}=0$
may have a nonzero one
$\delta^{\star}\ne 0$
at moderate and weak fields
(i.e. $\vert\Omega_0\vert<2\vert{\cal{I}}\vert$).
This nontrivial
$\delta^{\star}$
comes from the equation that follows from (4)
\begin{eqnarray}
\frac{\pi\alpha}{\vert{\cal{I}}\vert}
=\frac{1}{1-\delta^2}
\left(
{\mbox{F}}(\psi,1-\delta^2)
-{\mbox{E}}(\psi,1-\delta^2)
\right)
\end{eqnarray}
where ${\mbox{F}}(\psi,a^2)\equiv\int_0^{\psi}d\phi/
\sqrt{1-a^2\sin^2\phi}$
is the elliptic integral of the first kind.

Consider the case $\Omega_0=0$.
Looking for a solution of Eq. (5) that satisfies the inequality
$\delta\ll 1$
(that is the limit interesting for applications)
one observes that the r.h.s. of Eq. (5) can be rewritten approximately as
$\int_0^{1}dx/\sqrt{\delta^2+x^2}$
($x=\cos\phi$)
that yields
$\delta^{\star}\sim\exp
\left(-\frac{\pi\alpha}{\vert{\cal{I}}\vert}\right)$.
{\em post priori} we conclude that the small dimerization parameters
$\delta^{\star}$
occur for hard lattices having large values of
$\frac{\alpha}{\vert{\cal{I}}\vert}$.
One also notes that the obtained result coincides with the one
reported
in
[2]
up to a renormalization of the effective interspin coupling
$\vert I\vert\rightarrow \vert{\cal{I}}\vert=\sqrt{I^2+D^2}$.
Thus the Dzyaloshinskii-Moriya interaction leads to increasing of
the dimerization parameter
$\delta^{\star}$ characterizing the dimerized phase.

Consider further the case
$0<\vert\Omega_0\vert<2\vert{\cal{I}}\vert$.
Varying $\delta$ in the r.h.s. of Eq. (5) from 0 to 1
one calculates a lattice parameter $\frac{\alpha}{\vert{\cal{I}}\vert}$
for which the taken value of $\delta$ realizes an extremum of
${\cal{E}}(\delta)$.
One immediately observes that for
$\frac{\vert\Omega_0\vert}{2\vert{\cal{I}}\vert}\le\delta\le 1$
the dependence
$\frac{\alpha}{\vert{\cal{I}}\vert}$
versus
$\delta$
remains as that in the absence of the field,
whereas for
$0\le\delta<\frac{\vert\Omega_0\vert}{2\vert{\cal{I}}\vert}$
the quantity
$\frac{\alpha}{\vert{\cal{I}}\vert}$
starts to decrease.
From this one concludes that for hard lattices
the field
$\frac{\vert\Omega_0\vert}{2\vert{\cal{I}}\vert}
=\exp\left(-\frac{\pi\alpha}{\vert{\cal{I}}\vert}\right)$
makes the dimerization
unstable against the uniform phase.
The latter relation tells us that the Dzyaloshinskii-Moriya interaction
increases the value of that field.

It is generally known
[1]
that the increasing of
the external field leads to a transition from the dimerized phase
to the incommensurate phase rather than to the uniform phase.
Evidently, the incommensurate phase cannot appear in the presented
treatment within the frames of the adopted ansatz for the
lattice distortions
$\delta_1\delta_2\delta_1\delta_2\ldots\;$,
$\delta_1+\delta_2=0$.
To clarify a possibility of more complicated distortions
the chains with longer periods should be examined.

Alternatively, we may also assume different
dependences on $\delta$ for the iso\-tro\-pic coupling
and the Dzyaloshinskii-Moriya coupling,
for example,
$\vert I_1\vert=\vert I\vert(1+\delta)$,
$\vert D_1\vert=\vert D\vert$,
$\vert I_2\vert=\vert I\vert(1-\delta)$,
$\vert D_2\vert=\vert D\vert$.
Supposing that
$\delta\ll 1$
after simple rescaling arguments
one finds that the dimerization parameter
$\delta^{\star}\sim \frac{{\cal{I}}^2}{I^2}
\exp\left(-\frac{\pi}{\vert{\cal{I}}\vert}
\frac{{\cal{I}}^4}{I^4}\alpha\right)$.
Thus, in such a case the Dzyaloshinskii-Moriya interaction leads
to a decreasing of the dimerization parameter
characterizing the dimerized phase.
The value of the field which destroys dimerization
$\frac{\vert\Omega_0\vert}{2\vert{\cal{I}}\vert}
=\exp\left(-\frac{\pi}{\vert{\cal{I}}\vert}
\frac{{\cal{I}}^4}{I^4}\alpha\right)$
decreases as well.

To conclude, we have analysed a stability of
the spin-$\frac{1}{2}$
transverse $XX$ chain with respect to dimerization in the presence of
the Dzyaloshinskii-Moriya interaction
calculating for this purpose
with the help of continued fractions
the ground state energy for an arbitrary
value of the dimerization parameter.
Assuming that the ratio of the Dzyaloshinskii-Moriya coupling to
the isotropic coupling does not depend on the dimerization parameter
we have found that the Dzyaloshinskii-Moriya interaction leads to an
increasing of the effective interspin coupling
and thus to
some quantitative changes,
i.e. to an increasing of the value of
the dimerization parameter
which characterizes the dimerized phase
and the value of the field which destroys the dimerized phase.
In the other limiting case
when the Dzyaloshinskii-Moriya coupling does not depend
on the dimerization parameter
it
has an opposite effect
leading to a decreasing of the value of the dimerization
parameter and the value of the field which destroys dimerization.
The obtained results are in agreement with some earlier studies
of the thermodynamic properties of
spin-$\frac{1}{2}$ transverse $XX$ chains with Dzyaloshinskii-Moriya
interaction
[10].
Finally, it is known that the Dzyaloshinskii-Moriya interaction may
lead to drastical changes in spin correlations
[11]
and a study
of a relation of these changes to the spin-Peierls
instability seems to be an interesting issue.

\vspace{5mm}

The present study was partly supported by the DFG
(projects 436 UKR 17/20/98 and Ri 615/6-1).
O. D. acknowledges the kind hospitality of the Magdeburg University
in the spring
of 1999 when the main part of the paper was done.
He is also indebted to Mrs. Olga Syska for continuous financial support.

\end{document}